# Embryologie de l'œil

Dr Sophie Creuzet (CNRS), Département Développement et Evolution, UMR-9197, Institut de Neurosciences Paris-Saclay, Gif-sur-Yvette

Dr Heather Etchevers (INSERM), Génétique Médicale et Génomique Fonctionnelle UMR_S910, Université Aix-Marseille, Marseille

---

Afin d'appréhender les mécanismes qui sous-tendent la physiologie de la vision, la connaissance des bases embryologiques du développement de l'œil et de ses annexes est un prérequis indispensable. La morphogénèse oculaire débute au cours de la quatrième semaine de vie embryonnaire, alors que l'ébauche oculaire s'individualise des diverticules latéraux du cerveau antérieur par des mouvements morphogénétiques complexes. Elle sollicite la contribution respective des divers feuillets de l'embryon, le neurectoderme, l'ectoderme de surface, le mésoderme et les cellules de la crête neurale, pour l'élaboration de ses différentes composantes. Les perturbations des interactions cellulaires et des mécanismes moléculaires mobilisés au cours de ces étapes critiques sont responsables d'anomalies congénitales variées. Nous évoquerons, à cet égard, les processus embryologiques dont les dérégulations sont à l'origine des malformations oculaires et qui font, plus particulièrement, l'objet de chapitres détaillés dans cet ouvrage.

## Rappel de quelques principes généraux d'embryologie

Au stade précoce du développement, dès deux semaines après la fécondation, l'embryon est composé de deux feuillets cellulaires superposés: l'un, sus-jacent, *l'épiblaste* ou ectoderme primitif, et l'autre, sous-jacent, *l'hypoblaste* ou endoderme primitif. Formés de cellules cohésives, ces feuillets épithéliaux forment dans leur zone de contact, le *disque embryonnaire*, à partir duquel se développe l'embryon. Au-delà, les feuillets se poursuivent séparément et délimitent les vésicules extra-embryonnaires, destinées à former les annexes extra-embryonnaires, à savoir, l'amnios, et le placenta, situées respectivement au-dessus et en dessous du disque embryonnaire.

> **L'orientation des axes embryonnaire**
>
> L'organisation spatiale et la dynamique des mouvements morphogénétiques impliquent de fixer des axes de référence anatomique internes à l'embryon. L'embryon humain présente l'avantage —par rapport à certains organismes modèles en embryologie, tels que la souris ou le poisson— de se développer dans le même plan de l'espace, au cours des périodes critiques conduisant de la gastrulation à la neurulation. Alors que l'embryon est encore diblastique, c'est-à-dire composé de deux feuillets, un plan d'organisation « dorso-ventrale» peut lui être assigné. L'acquisition de cette polarité repose sur le fait que l'épiblaste, à l'origine du système nerveux central, se développe au pôle dorsal de celui-ci alors que l'hypoblaste, destiné à tapisser la face interne de cavités viscérales, détermine le pôle ventral. Lorsque la gastrulation s'engage, l'invagination des cellules d'épiblaste en position intermédiaire, conduit à la formation d'un troisième feuillet embryonnaire, le mésoderme. Cette organisation triblastique mobilise les cellules de l'épiblaste qui convergent vers une ligne longitudinale médiane, la ligne primitive où elles subissent un double changement : d'une part, elles cessent d'être ectodermiques pour s'intercaler entre l'épiblaste et l'hypoblaste, et deviennent mésodermiques ; d'autre part, elles adoptent une organisation tissulaire moins dense et cohésive que celle du feuillet épithélial, le mésenchyme, propice à des remodelages tissulaires rapides et des migrations à distance.
>
> La formation de la ligne primitive confère le plan de symétrie bilatérale qui sépare les côtés droit et gauche de l'embryon, et donne également les repères d'un axe « médio-latéral ». Chez tous les organismes bilatériens, la morphogenèse de l'embryon débute au pôle céphalique, qui définit la partie antérieure ou rostrale de celui-ci, puis gagne de proche en proche les niveaux plus postérieurs ou caudaux. Par conséquent, ce gradient de développement permet de définir précocement l'axe « antéro-postérieur » ou « rostro-caudal » de l'embryon.

Dans les jours qui suivent, un sillon médian se forme à la surface de l'épiblaste, dont la position définit l'extrémité caudale l'embryon. Ce sillon s'appelle la *ligne primitive* et marque l'endroit où se déroule la gastrulation (Figure 1A,B). Pour reprendre l'aphorisme du grand embryologiste Lewis Wolpert, le moment le plus important de la vie n'est ni la naissance, ni le mariage, ni la mort, mais la gastrulation [1]. A partir de ce stade, l'épiblaste du disque embryonnaire proprement dit, devient l'ectoderme définitif à l'origine du tissu épithélial qui tapisse et couvre la face externe de l'organisme. Lorsque la ligne primitive est à son extension maximale, les cellules les plus rostrales situées à sa base s'invaginent et s'agrègent en un amas de cellules mésodermiques axiales (Figure 1B, C). La ligne primitive engage ensuite une régression rostro-caudale, relative par rapport à l'allongement antérieur de l'embryon, au cours de laquelle, elle dépose dans son sillage le matériel cellulaire destiné à former la notochorde. De façon concomitante, la plaque neurale est induite dans l'ectoderme médial sus-jacent). Bien qu'initialement formée d'une seule couche de cellules, la plaque neurale se caractérise par un rapide épaississement, qui conduit à la spécification du neuro-ectoderme, et à sa démarcation de l'ectoderme latéral. La plaque neurale subit une réorganisation cellulaire et des mouvements complexes d'extension et de convergence, qui précédent la formation d'une gouttière neurale (Figure 1D). Les bords latéraux de cette gouttière se rejoignent progressivement pour fusionner le long de la ligne médiane dorsale. La fusion des bords du neuro-ectoderme permet, d'une part, de restaurer la continuité de l'ectoderme superficiel, destiné à former l'épiderme, et, d'autre part, d'internaliser le tube neural à l'origine de l'ensemble du système nerveux central (Figure 1E).

La fermeture du tube neural débute au niveau du futur cerveau moyen, puis gagne de façon bidirectionnelle, les niveaux plus rostraux et plus caudaux. En amont, le mécanisme laisse un neuropore antérieur qui se résorbe dans les jours qui suivent (Figure 1C, H). Les anomalies du développement qui surviennent au cours de ce processus de fermeture génèrent des malformations extrêmement sévères, telles que l'anencéphalie, qui ne sont pas compatibles avec la vie postnatale. Celles-ci peuvent être facilement et précocement décelées par échographie. La désorganisation extrême qui en résulte dans le tissu cérébral, a pour conséquence des remodelages importants et délétères des champs optiques, du fait de leur proximité.

Outre l'implication successive de l'ectoderme, du neuro-ectoderme, et, dans une moindre mesure, du mésoderme, l'ontogénèse de l'œil mobilise une ultime population cellulaire qui contribue de façon essentielle à la morphogénèse, l'organogénèse, et la physiologie optique : la *crête neurale*. Il s'agit d'une population de cellules qui a pour origine les bourrelets neuraux qui délimitent latéralement la gouttière neurale (Figure 2A). Avant la fermeture du tube, ces cellules sont épithéliales et liées au neuro-ectoderme, mais, à mesure que la fermeture du tube neural s'engage, elles se détachent des bourrelets latéraux et deviennent mésenchymateuses (Figure 2A, B). Leur individualisation s'opère selon une cinétique bidirectionnelle qui suit la fermeture du tube neural. Bien que ce processus soit très conservé chez les vertébrés, des variations subtiles peuvent exister dans la dynamique de leur délamination selon les espèces.

La crête neurale est une grande innovation qui a marqué l'histoire des chordés et constitue une caractéristique exclusive les vertébrés. Du fait de son caractère hautement multipotent, elle est considérée comme le quatrième feuillet germinatif de ce groupe phylogénétique. Son apparition au cours de l'évolution a permis l'acquisition d'une grande variété de caractères propres, parmi lesquels la formation d'un squelette crâniofacial comprenant les mâchoires, la face supérieure et le crâne. En outre, l'émergence de ses structures squelettiques a coïncidé avec l'accroissement et la sophistication du cerveau antérieur et des organes de sens.

Il est plus approprié de parler de « cellules de crête neurale » que de « crêtes neurales », qui désignent spécifiquement les bords de la gouttière neurale en cours de fermeture. Les cellules de la crête neurale (CCN) lorsqu'elles se détachent des bourrelets neuraux, démarrent d'importantes migrations qui les conduisent à essaimer dans tout l'embryon où elles se différencient en une remarquable variété de lignages et de dérivés [2]. Les dérivés de CCN des niveaux céphaliques et troncaux sont présentés dans la Figure 2C.

Outre une contribution particulièrement riche à l'ontogenèse, la crête neurale subsiste également chez l'adulte à l'état indifférencié au niveau céphalique, dans certains foyers, qui se comportent comme autant de réservoirs, ou « niches » de cellules souches, susceptibles de participer à des processus régénératifs variés [3]. Du fait de leurs capacités de différenciation plus étendues par rapport à celles du mésoderme, les cellules souches de la crête neurale font l'objet d'intenses recherches visant à maîtriser les conditions de leur utilisation pour l'ingénierie tissulaire et la médecine régénérative. Ceci est particulièrement le cas en ce qui concerne la cornée.

## Morphogenèse de l'œil au cours du deuxième mois de gestation

Le développement de l'œil proprement dit débute à 22 jours de gestation (J22), alors que la taille de l'embryon humain atteint 2 mm de longueur. Au niveau céphalique, tandis que la plaque neurale commence à se replier pour former le tube neural, des dépressions ou diverticules apparaissent à la face interne de la plaque, et marquent des évaginations latérales du neuro-ectoderme vers l'ectoderme de surface (Figure 3A, C, D). A ce niveau, la partie médiale de la plaque neurale est destinée à former une division majeure du cerveau antérieur, le diencéphale, à partir duquel se forment d'autres structures telles que l'hypothalamus et le chiasma des nerfs optiques, pour une distribution des axones indispensable la vision binoculaire. Une seconde division majeure se forme dans la partie latérale de la plaque neurale antérieure : il s'agit du télencéphale, à l'origine des hémisphères cérébraux qui vont croître en avant des diverticules optiques (Figure 3B, E).

Dans les jours qui suivent, alors que des unités métamériques de mésoderme troncal, les *somites*, se ségrégent de part et d'autre du tube neural en suivant l'élongation du corps (Figure 1F, 2G), les vésicules optiques issues des évaginations du neuro-ectoderme, s'élargissent (Figure 2B, 3E). La progression des vésicules optiques s'opère en direction de l'ectoderme de surface au contact duquel le neuro-ectoderme s'épaissit et détermine le disque rétinien vers J27. Leur croissance latérale est accompagnée par un afflux de cellules mésenchymateuses (Figure 4).

De façon réciproque, l'ectoderme de surface subit également une différenciation qui débute, là encore, par l'épaississement des cellules à son niveau (Figure 4B). Cet épaississement délimite la placode cristallinienne qui secondairement s'invagine jusqu'à former une vésicule cristallinienne (Figure 4B, C), puis s'individualise totalement de l'ectoderme de surface adjacent pour aboutir à la formation d'une lentille internalisée sous l'ectoderme, le *cristallin*. La formation de la placode cristallinienne est l'un des exemples les plus classiques du processus d'induction en biologie du développement, mettant en jeu des signaux morphogénétiques produits par le neuro-ectoderme et l'ectoderme de surface, et sollicitant également d'autres tissus, situés plus à distance tels que le mésoderme cardiaque et l'endoderme pharyngien.

La formation de la placode cristallinienne coïncide avec l'apparition d'une constriction à la face proximale de la vésicule optique, au niveau de son point d'attache à la paroi latérale du cerveau antérieur. Cette constriction, la *tige optique*, s'allonge et s'accentue au cours de la croissance à mesure que la morphogenèse de la vésicule optique gagne en sophistication. La lumière de la tige optique maintient une continuité entre la cavité de la vésicule optique, qui donne l'espace sous-rétinien, et le troisième ventricule, vésicule unique et médiale du diencéphale (Figure 4C).

A la fin de la quatrième semaine de développement, la vésicule optique est globalement sphéroïde et composée d'une monocouche de cellules. Elle subit ensuite une invagination spectaculaire par le biais d'élongations et de mitoses cellulaires qui accroissent la surface de tissu neuro-épithélial à son niveau, mais également par des changements cytosquelettiques, et de phénomènes de repliement qui aboutissent à la formation de la cupule optique.

Le disque rétinien situé initialement à l'apex de la vésicule (Figure 4C), est transitoirement superposé à la placode du cristallin : ces deux couches cellulaires d'origine distinctes sont liées par des pontages cellulaires temporaires. L'accroissement de la cupule optique n'étant pas uniforme à sa circonférence, une croissance différentielle conduit à la formation d'un sillon le long de la face distale et ventrale, dont les bords convergent pour former la fissure optique. A J29, deux invaginations concomitantes - du disque de la rétine et de la placode du cristallin - sont presque achevées (Figure 4D). Superficiellement, une petite dépression peut être observée alors que la lentille du cristallin est en cours d'internalisation. Autour de ce point, le territoire où l'ectoderme de surface tend à recouvrer son intégrité, est à l'origine de la future cornée.

La vésicule du cristallin se sépare définitivement de l'ectoderme de surface avant J36. Les cellules épithéliales du cristallin se referment sur une cavité et sont bordées extérieurement par une lame basale qui forme la capsule du cristallin. Au niveau de la fissure optique, le sillon longitudinal s'étend de la tige optique jusqu'à la cupule, qui parallèlement, s'élargit et s'invagine. Ce mouvement morphogénétique aboutit à la juxtaposition de la paroi distale et de la paroi proximale de la tige optique. Dans la fissure, une branche de l'artère ophtalmique, l'artère hyaloïde, et des cellules dérivées de la crête neurale (CCN) se trouvent incorporées à l'espace lentorétinal. A la fin de la sixième semaine de développement (6 sd), soit approximativement 8 semaines d'aménorrhée, les bords de la fissure se rejoignent et fusionnent en isolant dans le centre de la tige optique les vaisseaux hyaloïdes et le mésenchyme associé, à l'origine de l'artère et du veine centrale de la rétine (Figure 5A). La fermeture de la fissure optique commence au milieu de la tige optique et continue simultanément dans une direction proximale (vers le cerveau) et distale (vers la rétine). La fusion de la fissure s'achève en marge de la cupule optique en ménageant un orifice à l'origine de la pupille (Figure 5B).

> **Le colobome**
>
> Le terme « colobome » signifie « mutilation» en grec. Cependant, dans la pratique clinique, il désigne un défaut congénital du quadrant inféronasal de l'œil, intéressant l'iris ; l'uvée et la rétine. Les colobomes sont généralement sporadiques et bilatéraux. Ils sont le résultat d'une absence totale ou partielle de la fermeture de la fissure optique qui conduit aux phénotypes rencontrés. Ainsi, les colobomes peuvent être antérieurs, postérieurs (rétine, choroïde, nerf optique) ou antéropostérieurs pour les formes les plus graves. Cette malformation a pour conséquence un défaut de l'induction et de la formation des tissus de l'uvée.
>
> • *Le colobome de l'iris* apparaît comme étant un défaut inféronasal affectant le stroma, le muscle lisse et l'épithélium pigmentaire à ce niveau.
>
> • *Le colobome de l'uvée* se caractérise par l'absence de procès ciliaires et une atrophie du muscle ciliaire. Ces structures, en condition physiologique, garantissent l'intégrité de la chambre antérieure de l'œil en assurant deux fonctions essentielles: mécanique d'une part, grâce aux fibres zonulaires qui permettent le maintien du cristallin par des ligaments suspenseurs des corps ciliaires, et physiologique d'autre part, puisque les procès ciliaires secrètent l'humeur aqueuse. Dans un contexte colobomateux, le cristallin adjacent est en retrait, en raison d'une hypoplasie ou une insuffisance des fibres zonulaires. Dans les colobomes syndromiques associés à des malformations complexes (telles que certaines trisomies), une différenciation anormale du mésenchyme dans l'espace rétrocristallinien, peut survenir avec la formation ectopique d'autres dérivés de CCN, tels que du tissu adipeux ou du cartilage (Figure 5C).
>
> • *Les colobomes chorio-rétiniens* impliquant la rétine peuvent présenter d'important déficits pour la fonction visuelle. A proximité du colobome, la prolifération du tissu neuroblastique rétinien peut conduire à la formation de rosettes. Un défaut d'induction ou de différenciation de l'épithélium rétinien pigmentaire dans la zone du colobome est souvent associé à l'absence de la membrane de Bruch et du tissu choroïdien, alors que la sclérotique sous-jacente paraît normale. Ces colobomes postérieurs et inféronasaux sont ceux rencontrés dans l'association syndromique connue sous l'acronyme CHARGE (**c**olobome, maladies cardiaques [**h**eart en anglais], **a**trésie des choanes, **r**etard de croissance et/ou de développement psychomoteur ; hypoplasie **g**énitale, malformation de l'oreille [*ear*]), et laissant suspecter des mutations du gène *CHD7*.

Après la séparation de la vésicule cristallinienne de l'ectoderme de surface, ce dernier se referme pour donner le futur épithélium cornéen (Figure 6A). De façon concomitante, à J39, une vague de cellules mésenchymateuses d'origine CCN migre massivement le long de la cupule optique, et directement sous l'ectoderme de surface (Figure 6C-E). Cette migration s'opère selon trois vagues successives. Les premières cellules à coloniser ce territoire, s'accumulent à proximité du cristallin : elles adoptent une morphologie pavimenteuse et développent des contacts apicolatéraux. Ces contacts organisent des jonctions intercellulaires continues qui aboutissent à la formation de l'endothélium cornéen et du trabeculum. Deux vagues successives de mésenchyme viennent secondairement élaborer d'abord le stroma de la cornée, ensuite le stroma de iris et le mésenchyme de l'angle irido-cornéen (Figure 6D). A noter que l'épithélium cornéen reste d'origine ectodermique (Figure 6E ; Creuzet et al. 2005; Cvekl & Ashery-Padan 2014).

À la fin de la période embryonnaire (à la fin de 8 sd), la rétine est clairement structurée en deux composantes majeures étroitement contigües. Extérieurement, la couche mince de la cupule optique forme *l'épithélium pigmentaire rétinien* (EPR) ; elle est doublée intérieurement d'une couche tissulaire beaucoup plus épaisse, destinée à former *la rétine neurale* (Figure 5B-D). Ces deux couches sont séparées par un espace sous-rétinien étroit, vestige de la cavité ventriculaire de la vésicule optique. Vers 5 sd, l'accumulation de la mélanine peut être déjà mise en évidence dans l'EPR. La rétine neurale débute une différenciation centrifuge à partir de la couche intérieure neuroblastique située près de la tige optique. Simultanément, la cavité du cristallin disparaît par l'allongement considérable des cellules postérieures qui sont disposées parallèlement et organisent ainsi les fibres primaires du cristallin (Figure 6).

Du mésenchyme, dérivé majoritairement des CCN mais associé à l'endothélium de capillaires d'origine mésodermique, se condense autour de la surface externe de la cupule optique [6]–[8]. La couche la plus interne de ce mésenchyme, intimement juxtaposée à la membrane basale de l'EPR, forme la membrane choroïde (lamina uvéocapillaire), un tissu conjonctif lâche et très vascularisé (Figures 6, 9). Elle est en continuité via la tige optique avec une membrane, histologiquement et fonctionnellement analogue, qui tapisse la face externe du cerveau antérieur, les méninges. En effet, la membrane choroïde est homologue dans son origine embryonnaire de la pie-mère et de l'arachnoïde, qui enveloppent le cerveau antérieur [7]. Par ailleurs, la couche externe du mésenchyme condensé autour de la cupule optique forme la sclérotique, en continuité avec la dure-mère qui enveloppe l'ensemble du cerveau antérieur. A ce titre, il est important de souligner que les cellules cartilagineuses dont la différenciation forme l'orbite squelettique de l'œil sont également issues des CCN [2]. Sur le plan ontogénique, l'œil et le cerveau antérieur bénéficient par conséquent d'un soutien vasculaire et squelettique dérivé des CCN ; plus caudalement, c'est le mésoderme qui fournit la majeure partie du mésenchyme vasculaire et squelettique annexé au système nerveux central et périphérique.

> **Syndrome de Sturge-Weber**
>
> Dans le syndrome de Sturge-Weber, l'association de déficits du réseau capillaire facial du derme périorbital (se manifestant par des « tâche de vin »), d'un glaucome ipsilatéral et de calcifications épileptogènes des leptoméninges corticales est due à une mutation stéréotypée et activatrice d'un gène codant pour un relais moléculaire de signal impliqué dans la prolifération et la différenciation cellulaire [9]. Il a été montré que cette mutation apparaît *de novo*, après la fécondation, dans la lignée des cellules endothéliales et se révèle dans le secteur des vaisseaux dont le muscle lisse est assuré par les CCN.

Pendant la huitième et dernière semaine de la période embryonnaire, les axones des cellules ganglionnaires de la rétine progressent vers la tige optique. Ces axones s'allongent à l'intérieur de la tige puis vers le cerveau, formant ainsi le nerf optique. Parallèlement, se forment les fibres secondaires du cristallin, ainsi que les sutures de la lentille et le corps vitré secondaire.

En résumé, à l'issue de la période embryonnaire, l'œil est composé de structures épithéliales composées d'une cupule à double couche et enserrant un cristallin dérivé de l'ectoderme de surface. La cupule comporte une couche interne de neuroectoderme, à l'origine de la future rétine, et une couche fine externe d'épithélium pigmentaire (rétinien, ou EPR), en continuité avec le nerf optique. L'œil comprend également un important contingent mésenchymateux périoculaire, lui-même constitué d'une couche externe dense, formant la majeure partie de la cornée et de la sclérotique, et d'une couche vasculaire lâche qui forme la choroïde, le stroma de l'iris et les corps ciliaires (Figures 6, 7). A ce stade, l'embryon humain fait 3 cm de longueur, et le diamètre du globe oculaire est compris entre 1,5 et 2,0 mm.

**Conservation fonctionnelle génétique au cours de l'évolution de l'œil**

Le chapitre VII.5 résume certains des gènes nécessaires au développement de l'œil chez plusieurs espèces. Les lecteurs intéressés par les modèles animaux destinés à l'étude du développement de l'œil et également par les mécanismes qui ont conduit à la diversification des fonctions de ces gènes au cours de l'évolution sont incités à consulter une édition spéciale du *International Journal of Developmental Biology* 2004; 48 (*eg*. Bailey et al. 2004).

**Malformations importantes de tube neural et de l'œil**

Des malformations du tube neural et des vésicules optiques peuvent se produire dans le premier mois de la vie embryonnaire (voir chapitre V.1.2). Elles englobent:

• ***L'anophtalmie primaire***, résultat d'un échec de la formation de la vésicule optique. Les orbites ne contiennent pas de tissu oculaire. Cependant, dans certaines formes d'anophtalmie, les muscles extra-oculaires issus du mésoderme, les tendons et tissus conjonctifs issus des CCN, ainsi que les glandes lacrymales issues de l'association de l'ectoderme et des CCN sont bien présents. La présence de ces dérivés atteste, par conséquent, d'un début d'induction des structures oculaires, qui ont secondairement subit une dégénérescence. Cette malformation très rare est associée principalement à des mutations des gènes codant les facteurs de transcription RAX, OTX2 et SOX2, lesquels peuvent conjointement engendrer un spectre de malformations syndromiques affectant le développement du cerveau antérieur.

• ***La nanophthalmie et la microphtalmie*** correspondent, l'une et l'autre, à un défaut de croissance de l'ébauche oculaire. Dans ce cas le développement initial est bien engagé, mais les structures oculaires cessent de croître prématurément, ce qui produit un œil rudimentaire et hypoplasique. L'œil qualifié de « microphtalme » est petit à la naissance mais contient des éléments reconnaissables tels qu'un cristallin, une membrane choroïde et une rétine.

• *La synophthalmie* correspond à la fusion des deux ébauches oculaires qui peut résulter soit d'une malformation, soit de processus inductifs défectueux, soit dépendre d'un défaut de différenciation du tissu mésenchymateux entre les vésicules optiques. Il est rare qu'un seul œil (cyclope) se forme par ce mécanisme : dans la plupart des cas, les deux cornées et deux cristallins sont bien individualisés, de même que les iris et corps ciliaires correspondants. Par contre, les structures crânio-faciales et les annexes oculaires médianes sont manquantes, si bien que la sclérotique de la ligne médiane et le tissu uvéal peuvent aussi être absents ; dans ce cas le nerf optique peut être simple ou double. Cette malformation peut être associée soit à une délétion du chromosome 18, soit à une **holoprosencéphalie** due à des mutations du gène *SHH*, codant pour un facteur de croissance essentiel à l'expansion de la population de CCN céphalique, ou de gènes codant pour les effecteurs intracellulaires de cette voie de signalisation.

• *L'oeil kystique congénital* se caractérise par le développement d'une structure kystique désorganisée qui empêche la morphogénèse et l'invagination du disque de la rétine.

## La neurogenèse et structuration de la rétine et l'épithélium pigmentaire

Le disque de la rétine est destiné à se différencier en neurones rétiniens, tandis que la couche fine la plus externe de la vésicule optique est destinée à former l'EPR. D'architecture et de fonctions distinctes, ces deux couches sont néanmoins en continuité dans un angle aigu au niveau de la chambre antérieure. La transition à ce niveau s'accompagne de la différenciation de structures hautement spécialisées dans la fonction optique: c'est là que se forment l'iris, le corps ciliaire et le bord de la pupille. En raison de l'invagination de la cupule optique, la partie apicale de la rétine neurale primitive vient s'adosser à la surface apicale de l'EPR, aux dépens de l'espace intra-rétinien.

A l'instar des cellules épendymaires qui couvrent les espaces ventriculaires du cerveau, les cellules qui tapissent les surfaces juxtaposées de la rétine neurale primitive et du futur EPR sont ciliées (Figure 8). La différenciation des cils de la rétine neurale revêt une importance physiologique fondamentale dans la maturation des cellules réceptrices, les cônes et les bâtonnets, et la transduction du stimulus lumineux à leur niveau. La présence de ces cils primaires sur les cellules de l'EPR est également indispensable. De nombreux gènes dont la mutation est responsable de la rétinite pigmentaire touchent à la formation des cils dans l'EPR. Parmi ceux-ci, le gène *GPCR* est impliqué dans la majorité des cas liés au chromosome X. La compréhension du rôle du cil primaire dans l'organisation et la signalisation épithéliale reste un sujet actuel de recherche.

De l'extérieur vers l'intérieur de la rétine mature, plusieurs couches histologiques deviennent identifiables (Figure 9). Elles comprennent six types de cellules nerveuses spécialisées :

- L'épithélium pigmentaire rétinien (EPR) ;
- La couche des segments externes (SE) des photorécepteurs : cônes (1) et bâtonnets (2) au contact de l'EPR ;
- La « membrane » limitante externe (mle) ;
- La couche nucléaire ou granulaire externe (CNE), contenant les corps des photorécepteurs;
- La couche plexiforme externe (CPE), où les dendrites des cellules radiaires bipolaires (3) et les dendrites des cellules horizontales (4) intègrent les signaux des photorécepteurs ;
- La couche nucléaire ou granulaire interne (CNI), comprenant les corps des interneurones bipolaires, horizontaux et amacrines (5) ;
- La couche plexiforme interne (CPI), contenant les axones de cellules bipolaires, connectés aux dendrites des cellules ganglionnaires, ainsi que les ramifications des cellules amacrines ;
- La couche des corps des cellules ganglionnaires (CG ; 6);
- La couche des axones ganglionnaires (AG) qui convergent vers le nerf optique ;
- La membrane limitante interne (mli).

La rétine neurale primitive, à l'origine de la majorité de la rétine mature, se compose tout d'abord d'une zone nucléaire et d'une zone acellulaire intérieure. La zone nucléaire correspond au neuroépithélium ventriculaire prolifératif du tube neural, et contient des cellules multipotentes. Les couches interne et externe de la cupule optique à ce stade ont des lames basales distinctes : celle de la couche intérieure donne la membrane limitante interne, et celle de la couche externe, la membrane de Bruch. La différenciation des couches neurales rétiniennes commence au pôle postérieur et progresse d'une manière centrifuge, donnant un gradient de différenciation de la rétine neurale à l'intérieur de l'œil autour de 7 sd. L'activité mitotique de la rétine neurale primitive est également plus grande dans la couche neuroblastique externe germinative. Les cellules nouvellement formées migrent vers l'intérieur de la cupule au niveau de la zone marginale pour donner la couche neuroblastique interne. Ces deux couches sont bien individualisées vers 6-7 sd par une zone cellulaire moins dense, mais également proliférative, connue sous le nom de la couche transitoire de Chievitz [11]. Des études de cartographie réalisées chez de nombreux modèles vertébrés ont montré la dynamique de ce processus en suivant la fluorescence émise par une protéine transgénique, produite de façon constitutive par les clones de cellules mosaïques [12]. Cette approche a montré la structuration en colonnes et le début de la stratification de la rétine, secondairement suivis par une ramification latérale.

Sur la face interne de la couche nucléaire, les axones des cellules ganglionnaires, premiers neurones à se différencier, convergent vers la tige optique. Une zone où les processus des cellules de la couche nucléaire intérieure s'entremêlent, la couche plexiforme interne, devient identifiable, au détriment de la couche transitoire de Chievitz qui disparaît (Figure 9). Les premières cellules différenciées de la couche neuroblastique interne donnent les cellules radiaires gliales de Müller, et les cellules amacrines, qui forment ainsi la couche nucléaire interne à partir du pôle postérieur de la rétine. Peu de temps après, les cellules bipolaires et horizontales se différencient dans la couche neuroblastique externe et migrent vers la nouvelle couche nucléaire interne. Les composants cellulaires restant de la couche neuroblastique extérieure forment ensuite la couche nucléaire externe, contenant les corps cellulaires des photorécepteurs (cônes et bâtonnets). La zone où les fibres de cette couche se mêlent à celles de la couche nucléaire interne constitue la nouvelle couche plexiforme externe (Figure 9). La

« membrane » limitante externe, n'est pas une membrane à proprement parler, mais se manifeste par l'alignement et la densité des jonctions serrées impliquant les photorécepteurs et les cellules de Müller.

Le développement oculaire est également marqué par d'autres étapes importantes pour l'élaboration de la rétine. Elles concernent, d'une part, la formation de la microglie (c'est-à-dire des macrophages tissulaires résidents) dont les cellules investissent la rétine via le système vasculaire rétinien et sous-rétinien (10-12 sd), et d'autre part, la synaptogenèse qui débute à partir des pédoncules des cônes à 4 mois de gestation, et à partir des sphérules des bâtonnets à 5 mois.

Un des événements frappants du développement de l'œil est l'apparition de la mélanine dans l'EPR embryonnaire vers J28. Initialement, l'EPR est un épithélium pseudostratifié cilié et mitotiquement actif. Les cils disparaissent alors que commence la mélanogénèse. Les cellules de l'EPR acquièrent une forme hexagonale et s'organisent en épithélium cubique simple, bien qu'une organisation pseudostratifiée se maintienne dans la rétine périphérique plus longtemps. Au cours du quatrième mois, l'EPR présente des microvillosités apicales, peu ou pas de replis basaux, des interdigitations basolatérales primitives, et des vésicules, *les mélanosomes*, qui séquestrent la mélanine dans le cytoplasme (Figure 10). L'activité mitotique a lieu très tôt dans le développement mais est, pour l'essentiel, terminée à la naissance. Par conséquent, la croissance de l'œil, et de l'EPR proprement dit, se fait par hypertrophie, c'est-à-dire par élargissement des cellules existantes. Les premières composantes de la membrane de Bruch, la lame basale de l'EPR, sont reconnaissables dès le stade de la cupule optique. Des fibrilles de collagène sont ensuite déposées sous la lame basale vers 10 sd ; la première ébauche de la couche élastique peut être détectée dès 15 sd, et devient fenestrée à la mi-gestation.

> **Malformations de la rétine**
>
> Des défauts dans l'organisation des cellules neuroblastiques de la rétine conduisent à l'épaississement et à la distorsion de l'architecture des réseaux neuronaux. Cette «dysplasie» rétinienne se manifeste par la formation de foyers de cellules neuroblastiques égarées au cours de la différenciation, formant alors des structures sphériques ou ovoïdes, appelées rosettes. Des anomalies plus subtiles peuvent survenir dans la formation de bâtonnets et les cônes qui peuvent conduire à des défauts de la vision des couleurs, ou à une mauvaise acuité visuelle associée à un nystagmus congénital.

## Maculogenèse et dépression fovéale. Spécialisation de la périphérie.

La maculogenèse se manifeste vers la mi-gestation par une augmentation localisée de la densité de cellules ganglionnaires situées au bord temporal du disque optique. A 6 mois, la couche des cellules ganglionnaires peut atteindre une profondeur de 8 à 9 cellules dans cette région. La couche nucléaire externe épaissie consiste principalement en des cônes immatures. Vers le 7ème mois, un déplacement des cellules ganglionnaires vient à former une dépression fovéale. Au 8ème mois, il n'y a plus que deux couches de cellules ganglionnaires à ce niveau et à la naissance la couche est monocellulaire. Pendant les 4 mois qui suivent la naissance, la couche de cellules ganglionnaires et la couche nucléaire interne se retirent jusqu'aux marges de la fovéa, ne laissant que des cônes dans la fovéa. L'allongement des segments internes et externes se poursuit au cours des mois suivants.

### Rétine périphérique

Jusqu'à 10-12 sd, la périphérie de la rétine ne s'étend pas jusqu'au bord de l'intérieur de la marge de la cupule optique. A 14 sd, elle se termine au niveau des futurs procès ciliaires, nouvellement formés. La *pars plana* définitive ainsi que l'*ora serrata* rudimentaire et la *pars plicata* sont présentes vers 6 mois de gestation. La *pars plana* et la région de l'*ora serrata* à l'équateur de l'œil continuent de s'étendre après la naissance avec la croissance continue du globe oculaire, qui se poursuit jusqu'à 2 ans d'âge. A la naissance, la zone de la rétine est approximativement de 600 mm$^2$ et atteint 800 mm$^2$ vers l'âge de 2 ans.

## Vascularisation de la rétine. Rétinopathie de la prématurité.

Issue de l'artère carotide interne, l'artère ophtalmique se ramifie pour donner l'artère hyaloïde qui s'incorpore à la fissure optique (Figures 5 à 7). L'artère hyaloïde, après avoir émergée de l'axe de la tige optique, se ramifie entre la surface du cristallin et la zone marginale de la rétine neurale primitive (espace lentoretinal). Avec la croissance de la cupule optique et la formation de la cavité vitréenne, l'artère hyaloïde traverse le corps vitré primitif, dans le canal hyaloïde ou *canal de Cloquet*, afin d'atteindre la surface postérieure du cristallin. La veine hyaloïde suit le même chemin en sens inverse.

Au début du 4ème mois de développement, des bourgeons angiogéniques se ramifient à partir des vaisseaux hyaloïdes sur le disque optique par division cellulaire et intercalation. Ces bourgeons sont constitués d'abord de cellules endothéliales, puis accompagnés par des péricytes d'origine CCN et des macrophages (Figure 6D). Les cordons initiaux de cellules endothéliales se canalisent et forment de nouveaux vaisseaux qui longent la couche de fibres nerveuses vers la rétine périphérique et progresse à une vitesse d'environ 0,1 mm par jour, pour atteindre l'*ora serrata* vers le huitième mois. En même temps, les bourgeons pénètrent également dans la profondeur de la rétine neurale jusqu'à la frontière extérieure de la couche nucléaire externe, selon une cinétique qui se poursuit après la naissance (Figure 11). A ce niveau, ils forment un réseau polygonal de vaisseaux, le plexus rétinien extérieur [13]. La partie de l'artère hyaloïde à l'intérieur de la rétine neurale donne l'artère centrale. Les capillaires se rejoignent et développent des jonctions serrées mais aussi communicantes immatures, cependant leurs lames basales sont incomplètes.

> **La rétinopathie des prématurés**
>
> L'hypoxie et l'hyperoxie sont susceptibles d'entraîner des effets stéréotypés sur la formation de réseaux capillaires de la rétine. En effet, les protéines de réponse au stress de l'hypoxie se fixent naturellement sur les promoteurs de gènes qui favorisent l'angiogenèse. A l'inverse, si des nourrissons prématurés sont placés dans un environnement où la pression d'oxygène est élevée, comme cela se pratiquait autrefois dans les couveuses, cela peut provoquer un retard ou une réduction de la vascularisation de la rétine par régression des microvaisseaux et l'inhibition de la formation de bourgeons vasculaires. Au retour à un taux et une pression d'oxygène normaux, les tissus subissent à nouveau une hypoxie localisée dans la rétine, ce qui entraîne des épisodes de néovascularization anormale au sein de la rétine et du vitré, connus cliniquement comme la rétinopathie des prématurés. Les grands prématurés nés à moins de 30 sd sont particulièrement à risque et peuvent dans une certaine mesure être traités par photocoagulation laser et, plus récemment, par des agents contre le VEGF (facteur de croissance de cellules endothéliales vasculaires). La rétinopathie des prématurés reste un grand problème de santé publique dans le monde [14], [15]

## Le nerf optique et la fissure choroïdienne

La tige optique assure, en son centre, la communication et la circulation des fluides entre la cavité du cerveau antérieur (futur troisième ventricule du cerveau) et la cavité des vésicules optiques en cours de développement (Figure 4B). Vers la fin de la 4$^e$ semaine de gestation, la tige est remplie de fluide et bordée par des cellules neuroectodermiques. L'invagination de la tige optique pour la formation de la fissure choroïdienne à sa face ventrale est concomitante de l'invagination de la vésicule optique en cupule optique. Pour la tige optique, ces mouvements morphogénétiques se traduisent par la formation d'une double couche de neuroectoderme qui se replie et s'affaisse sur elle-même : ce collapsus entraîne la disparition de la cavité intermédiaire de fluide, et conduit à l'incorporation des vaisseaux hyaloïdes et le mésenchyme d'origine CCN qui les accompagne (Figure 4). Les bords de la tige optique se ferment d'abord sur les vaisseaux hyaloïdes près du cerveau vers 5-6 sd, puis leur fusion s'étend distalement pour atteindre le bord inférieur de la cupule. Par la croissance asymétrique de la rétine temporale, le nerf est déporté du côté nasal vers la fin du premier trimestre. Chez le fœtus, les tiges optiques se situent à environ 65° par rapport au plan mi-sagittal, alors que par croissance différentielle, le nerf optique qui en est issu se situe plutôt à 40° chez l'adulte.

La tige optique conduit des axones des neurones ganglionnaires de la rétine vers le cerveau. Le neurectoderme extérieur de la tige se différencie en cellules gliales qui entourent le nerf optique mais donnent aussi la composante gliale de la lame criblée [16]. A l'intérieur du nerf optique, les faisceaux axonaux sont entourés par d'autres cellules gliales myélinisantes qui se sont différenciées à partir de la couche interne de la tige optique [17]. Comme pour le reste du diencéphale, le mésenchyme dérivé des CCN adjacent fournit les composantes conjonctives et méningées du nerf optique.

### Malformations de la tête du nerf optique

Lorsqu'un défaut de la fermeture de la fissure optique survient dans sa partie postérieure, il peut être à l'origine d'un colobome de la tête du nerf optique. Les colobomes touchant le nerf optique peuvent être isolés ou associés à des colobomes chorio-rétiniens. Situées dans la partie inféronasale, ces malformations peuvent être associées à un staphylome, c'est-à-dire une hernie de la sclérotique. De même, elles peuvent impliquer une hernie rétinienne dans les méninges du nerf optique, parfois sous la forme d'une simple fosse localisée au bord du disque. L'appellation historique de « morning glory syndrome » qu'on retrouve parfois dans la littérature est due à l'aspect qu'évoque son observation en fond d'œil et sa ressemblance avec la fleur de liseron ou ipomée.

L'importance clinique de ces manifestations tient aux complications qu'elles peuvent provoquer, et notamment l'infiltration de fluide sous la macula, susceptible d'entraîner le décollement de cette dernière. D'autres formes présentent une excavation très importante de la papille et une leucocorie évidente. Du tissu adipeux, du muscle lisse ou des cellules gliales ectopiques peuvent être présents à l'intérieur des méninges, ainsi que des hernies correspondantes du cerveau [18]. Les formes sévères et syndromiques peuvent s'accompagner d'agénésie du corps calleux ou des malformations de l'axe hypothalamo-hypophysaire [19].

## Mise en place de la cornée

L'ectoderme de surface qui recouvre son intégrité après individualisation de la vésicule cristallinienne est destiné à former l'*épithélium cornéen*, stratifié par quelques couches cellulaires. Une fois l'épithélium cornéen formé, vers 5 sd, les CCN du mésenchyme périoculaire migrent le long du bord de la cupule optique, pour investir l'espace compris entre l'ectoderme et la surface antérieure du cristallin, et constituer une couche oligocellulaire, l'*endothélium cornéen*. Cet endothélium lui-même repose sur une lame de matrice basale qui préfigure la *membrane de Descemet*. Vers 7 sd, une deuxième vague de CCN vient s'infiltrer entre l'endothélium et l'épithélium cornéen pour former le stroma cornéen, au départ un mésenchyme lâche, mais qui, à mesure que ses cellules se différencient, tend à se stratifier (Figures 6, 11).

Dans les semaines qui suivent, les paupières se forment à partir d'une expansion de l'ectoderme périoculaire, soutenue par l'accumulation du mésenchyme sous-jacent d'origine crête neurale. Ensuite, les paupières se soudent vers 9-10 sd et restent fermées pendant plusieurs mois (Figure 11). Vers la mi-gestation, toutes les couches de la cornée sont présentes, à l'exception de la « membrane » de Bowman, qui est une couche épaisse de collagène acellulaire sous l'épithélium cornéen. Les faisceaux de collagène dans le stroma s'organisent en lamelles fasciculées, et les cellules stromales fibroblastiques, alors désignées comme des *kératocytes*, s'aplatissent et développent une intense activité sécrétrice, de protéoglycanes notamment. La cornée reçoit un important contingent d'axones sensoriels provenant du ganglion trigéminal (Figure 12 A, B), qui intéresse d'abord la périphérie de la cornée, mais qui s'étend de façon centripète entre le $3^e$ et $5^e$ mois [20]. Le processus de maturation de la cornée est initié dans les couches les plus profondes de la cornée, puis progresse plus superficiellement. Une vague de différentiation des CCN, qui constituent la majeure partie de la cornée, les nerfs sensoriels et la sclère, précède cette maturation [21]. La composition des glycosaminoglycanes, tel que l'acide hyaluronique, change au cours du temps. Cette fluctuation permet de favoriser l'hydratation et le gonflement de la gelée matricielle pendant la période où les paupières sont fusionnées, puis sa compaction à l'ouverture de ces dernières, à partir de la 24$^e$ semaine [22]. Ces changements s'accompagnent d'un arrêt quasi complet de la prolifération cellulaire au sein de la cornée, la vie durant.

> ### Malformations congénitales de la cornée
>
> Lorsque la membrane basale qui sous-tend l'endothélium cornéen, la membrane de Descemet, présente à sa périphérie —désignée par la ligne ou l'anneau de Schwalbe— un épaississement anormal au point qu'il devienne proéminent et ectopique, on parle *d'embryotoxon* (du grec « *toxon* », un arc). Cette malformation est visible par l'examen de l'angle irido-cornéen en gonioscopie, et affecte près d'une personne sur six dans la population. L'anomalie d'Axenfeld regroupe l'embryotoxon postérieur avec des adhésions, ou synéchies irido-cornéennes périphériques. Si celles-ci sont associées à un glaucome et/ou d'autres malformations variables telles que la microcornée, l'hypoplasie de l'iris, la polycorie (pupilles multiples), la corectopie (pupille excentré), des dysmorphies faciales (hypoplasie maxillaire, malformations dentaires), il s'agit alors du syndrome d'Axenfeld-Rieger. La combinaison d'adhésions iriennes à la face postérieure de la cornée et de l'opacification cornéenne correspond à l'anomalie de Peters, souvent étroitement associée au développement d'un glaucome (voir chapitre V.7). Ces malformations sont génétiquement hétérogènes : elles font partie d'un spectre de mutations qui affectent l'activité des gènes codant pour des facteurs de transcription nécessaires au développement précoce de l'œil chez tous les vertébrés (chapitre VI.5).

## Développement du cristallin

La placode cristallinienne s'épaissit dans l'ectoderme en regard des vésicules optiques vers J27. La différenciation du cristallin requiert la mise en jeu de deux types de signaux protéiques échangés entre les contingents cellulaires présents, ainsi que la compétence de la placode à y répondre : d'une part, un signal inductif, produit par le neuroepithélium, et d'autre part, un signal répressif du mésenchyme d'origine crête neurale, qui permet de circonscrire la formation de cristallin à un endroit précis, en inhibant et contrecarrant le potentiel cristallinien de l'ectoderme environnant.

Après l'induction de la placode, l'ectoderme cristallinien s'invagine en se renfermant sur lui-même, afin de former une vésicule creuse. L'individualisation du cristallin de l'ectoderme de surface vers J33 marque le moment où la chambre antérieure commence à se façonner. A la face postérieure de la vésicule cristallinienne, les cellules organisées en couche simple tendent à se différencier sous l'effet inducteur de la rétine [23], et débutent une élongation qui les conduit à croître vers la lumière de la vésicule cristallinienne et en direction de l'ectoderme (Figure 7) ; cette étape d'élongation est indispensable à l'acquisition du pouvoir réfractif du cristallin [5].

> **Malformations du cristallin**
>
> La juxtaposition de la vésicule optique et de l'ectoderme compétent est absolument requise pour permettre l'induction d'un cristallin et son bon positionnement. Les perturbations précoces de ce processus conduisent à l'aphakie primaire congénitale (du grec « *phakos* », lentille), c'est-à-dire, l'absence de cristallin. De même, la taille de la placode cristallinienne dépend du contact initial entre la vésicule optique et l'ectoderme. Ainsi, une petite placode amènera la microphakie. Toutefois, ces malformations surviennent rarement de façon isolée et sont souvent associées à des anomalies de la chambre antérieure telles que des dysgénésies du segment antérieur.
>
> Plusieurs malformations sont le résultat d'une séparation défectueuse de ces tissus à des stades plus tardifs. A titre d'exemple, il s'agit :
>
> - des adhésions kératolenticulaires typiques de l'anomalie de Peters
> - du lenticône, qui se manifeste par une protrusion antérieure du cristallin provoquée par un manque de consistance de sa membrane basale, dû à l'absence d'un de ses composants en collagène, notamment dans le syndrome d'Alport
> - de la cataracte sous-capsulaire antérieure provoquée par une opacification de cette membrane basale qui s'épaissit dans la partie antérieure, au cours du temps.

## Mise en place du vitré

Le *vasa hyaloidea* et la tunique vasculaire du cristallin (ou *tunica vasculosa lentis*) désignent un abondant réseau capillaire issu de l'artère hyaloïde qui pénètre l'espace du vitré primitif à travers la fissure choroïde et la tige optique, pour gagner la face postérieure et latérale du cristallin. A leur niveau, les vaisseaux sont dotés d'un endothélium non-fenestré accompagné d'une couche périvasculaire simple adossée à une membrane basale. Ce réseau capillaire hyaloïde destiné à assurer la vascularisation du cristallin n'a qu'une existence temporaire, le temps que se façonnent les corps ciliaires et le *canal de Schlemm* dans la chambre antérieure. Le vasa hyaloidea et la tunique vasculaire du cristallin disparaissent par un processus physiologique de thrombose qui laisse alors le corps vitré secondaire ou définitif, hyalin, avasculaire et acellulaire.

> **Malformations de vitré**
>
> La persistance anormale de la tunique vasculaire du cristallin est responsable d'une malformation congénitale considérée comme une persistance du vitré primitif [24]. Cette anomalie, généralement unilatérale, entraîne l'opacification du cristallin, et peut être associée à une élévation de la pression intraoculaire, ainsi qu'à une microphtalmie. La masse de tissu fibreux/glial au niveau de la tête du nerf optique est connu sous le nom de *papille de Bergmeister* et représente le vestige glial des vaisseaux hyaloïdes incomplètement atrophiés.

## Morphogenèse de l'uvée

L'uvée désigne un complexe vasculo-pigmentaire situé en position intermédiaire dans l'œil. Elle forme un continuum structural et fonctionnel qui s'étend de la partie antérieure à la région postérieure, en comprenant l'iris et les corps ciliaires (uvée antérieure) et la membrane choroïde (uvée postérieure). Cette structure assure par son réseau capillaire une fonction de soutien métabolique et de nutrition pour l'iris et les corps ciliaires. Elle participe également à la fonction visuelle par l'absorption et la limitation de la réflexion lumineuse sur la rétine ce qui favorise un bon contraste visuel.

## La choroïde

Comme nous l'avons déjà évoqué, la membrane choroïde est une toile vasculaire qui épouse la face externe de la rétine pigmentaire (Figure 10). Elle est juxtaposée à cette dernière par la membrane de Bruch qui forme à ce niveau une lame basale riche en collagène et en fibres élastiques. La choroïde est classiquement décrite comme une succession de couches concentriques, dont la membrane de Bruch constitue la strate la plus interne. Extérieurement, elle est limitée par la *lamina suprachoroidea*, composé d'un réseau de fibres élastiques avasculaires, mais riches en mélanocytes. Encadrés par ces deux membranes, le réseau vasculaire du stroma de la choroïde présente une organisation topographique coaxiale selon le diamètre de ses vaisseaux, allant de la choriocapillaris—les capillaires choroïdes— en face interne, à la couche de Slatter, en position intermédiaire et composée de vaisseaux de taille moyenne, puis la couche de Haller, plus périphérique et formée de vaisseaux de plus large diamètre. La choroïde est absolument requise pour l'équilibre homéostasique des structures auxquelles elle est adossée : la rétine pigmentaire et la sclérotique, postérieurement, et l'iris et les corps ciliaires, antérieurement. Elle reçoit un soutien essentiel des CCN qui forment les péricytes qui doublent l'endothélium de ces vaisseaux ainsi que les cellules pigmentaires qui les accompagnent.

## Le corps ciliaire

Le développement du corps ciliaire présente des similitudes avec le développement de l'iris (Figure 6). Il s'agit de la juxtaposition de l'épithélium pigmenté, auquel sont appendues les fibres zonulaires qui sous-tendent le cristallin, et du tissu musculo-conjonctif dérivé des CCN, qui sont impliquées dans la production et la sécrétion de l'humeur aqueuse. L'épithélium ciliaire est marqué par 70-75 replis qui assurent l'insertion les fibres zonulaires. La production aqueuse débute dès 20 semaines, et coïncide avec des changements concomitants dans l'angle irido-cornéen.

Le muscle ciliaire, formé des fibres musculaires lisses orientées longitudinalement, se termine dans la région trabéculaire. Les fibres du muscle ciliaire circulaires ou radiales se différencient beaucoup plus tard au cours du développement, puisque leur formation n'est pas toujours totalement achevée jusqu'à environ 1 an d'âge.

### L'iris

L'iris constitue le diaphragme de la pupille. Il comporte les muscles lisses de l'iris, le sphincter, qui rétrécit la pupille, et les deux muscles dilatateurs de la pupille. Il s'agit de structures remarquables du fait de leur origine embryologique, seuls exemples dans l'organisme où les structures musculaires striés dérivent des CCN et non pas du mésoderme. Les muscles iriens, d'action antagoniste, reçoivent une innervation parasympathique cholinergique pour le sphincter, et sympathique adrénergique, pour les muscles dilatateurs, dont la maturation s'opère en fin de gestation.

Un abondant tissu conjonctif est associé aux muscles iriens. Dérivé comme ces derniers des CCN, il renferme les mélanocytes responsables de la couleur de l'iris. La pigmentation de l'iris est initiée au cours du $4^e$ mois de gestation et s'opère de la périphérie vers le centre, jusqu'à la naissance et dans les mois qui suivent. Il peut cependant encore évoluer pendant les premières années en fonction de l'épaisseur du stroma. Les yeux foncés laissent passer la coloration de l'épithélium du fond de l'iris à la naissance, pour céder la place aux mélanocytes étoilés pigmentés au sein du stroma. En revanche, la couleur des yeux clairs résulte de l'interférence et de la réflexion de la lumière directement sur les fibres de collagène du stroma moins pigmenté.

> **Aniridie**
>
> L'aniridie, rarement isolée, est généralement associée à une malformation de la chambre antérieure et plus particulièrement de l'angle irido-cornéen. Cette malformation correspond à un iris histologiquement anormal et hypoplasique avec stroma hypercellulaire, souvent associé à une prolifération aberrante de l'épithélium pigmentaire, combinée à une anomalie ou une hypoplasie du système d'écoulement. Cette anomalie coïncide également avec des opacités ou des ectopies du cristallin et une hypoplasie du nerf optique. Compte tenu de l'importance développementale et physiologique du rôle qu'exerce la crête neurale à ce niveau, cette malformation est considérée comme une neurocristopathie, c'est-à-dire une pathologie propre aux CCN. Cette malformation génétique est le résultat de mutations dans un gène codant *PAX6*, un facteur de transcription particulièrement important au fil de l'évolution de l'œil.

## Angle de la chambre antérieure et écoulement des liquides

Une autre particularité des compétences de la crête neurale à l'angle irido-cornéen concerne la formation du canal de Schlemm, qui assure le drainage de l'humeur aqueuse et est, par conséquent, essentiel à l'homéostasie de la chambre antérieure. Cette structure est composée d'un épithélium simple et contractile, exclusivement formé des CCN. Elle forme l'extrémité « aveugle » d'un vaisseau de type lymphatique, qui par l'activité pulsatile de son endothélium draine l'humeur aqueuse vers les veines aqueuses et épisclérales. De cette manière, le canal de Schlemm maintient une pression constante dans la chambre antérieure. Les altérations de la capacité de résorption du liquide par ce canal entraînent un glaucome.

> **Glaucome congénital**
>
> Un glaucome congénital peut être généré par la présence obstructive ou la persistance anormale de la « membrane » de Barkan. Il ne s'agit pas d'une membrane, proprement dit, mais d'un artefact. Une couche apparemment continue est en fait un artéfact histologique des cellules endothéliales qui se forme à la surface du réseau trabéculaire [25]. Au cours du développement, pendant le 4ème mois de gestation, ces cellules ne couvrent qu'un tiers voire la moitié de la région trabéculaire et sont déjà discontinues pour permettre une communication entre la chambre antérieure fœtale et les espaces intertrabéculaires. Cependant, dans la situation pathologique où cette membrane gagne l'angle irido-cornéen et persiste à ce niveau, elle fait obstacle à la résorption du l'humeur aqueuse, et est susceptible d'entraîner un œdème de la cornée. Sa présence pathologique et ectopique dans la mégalocornée nécessite une chirurgie filtrante en urgence.

## Annexes palpébrales : les paupières et glandes lacrymales

Il est important de garder à l'esprit, qu'à la face antérieure de l'œil, l'ectoderme situé à la périphérie du territoire dévolu à la formation de l'épithélium cornéen, est impliqué dans la formation d'annexes périoculaires non moins essentielles à la protection et à l'homéostasie de l'œil. Il s'agit du complexe palpébral qui comprend des enveloppes protectrices motiles, les paupières, ainsi que les glandes qui secrètent le film lacrymal à la surface du globe oculaire. Cet ensemble est façonné, au cours du $2^{ème}$ et du $3^{ème}$ mois de gestation, par des interactions multiples entre l'ectoderme de surface et le mésenchyme sous-jacent issu des CCN.

Les paupières naissent d'un repli de l'ectoderme et apparaissent telles des excroissances paires qui encadrent la partie supérieure et inférieure de la cornée (Figure 12C-E). Les CCN forment à ce niveau la musculature striée des muscles orbiculaires, droits, et releveurs de la paupière, ainsi que les muscles de Müller et la composante lisse du plan fibro-élastique. Les myofibres issues de la différenciation des CCN sont dotées d'une innervation réflexe, mais aussi volontaire. Quant au complexe glandulaire des paupières (Figure 12F-H), il se forme par de multiples invaginations de l'ectoderme qui génère un ensemble de glandes lacrymales vers la fin du $4^e$ mois de gestation [26]. Il s'agit notamment des glandes lacrymales principales qui s'ouvrent à la surface du tissu conjonctival et du cul de sac et sont responsables de larmes reflexes qui concourent à maintenir l'asepsie et l'hydratation de la surface oculaire. Quant aux larmes basales, elles sont produites par les glandes lacrymales accessoires, qui chacune contribue de façon variée à la composition et au renouvellement du film lacrymal. Les glandes lacrymales accessoires tapissent le tissu conjonctival et sont responsables des sécrétions muciniques qui composent la couche profonde du film lacrymal, directement à la surface de la cornée. Les glandes sébacées de Meibomius se développent le long du tarse palpébral, à la face interne de la paupière, et s'ouvrent à la face interne pour déposer une sécrétion aqueuse intermédiaire. Les glandes sudoripares de Moll sont situées à l'apex de la paupière et à proximité du cil, dont la sécrétion lipidique forme la couche la plus superficielle du film lacrymal.

## Vers une classification embryologique des malformations congénitales

La classification des malformations congénitales ou anomalies du développement est difficile pour plusieurs raisons. Tout d'abord, l'étiologie est souvent inconnue, même quand une cause génétique ou environnementale unique est suspectée. Il est par conséquent souvent difficile d'attribuer une responsabilité exclusive de ces agents ou des événements dans ces processus pathologiques, notamment parce que les causes sont souvent multifactorielles, et mettent en jeu des facteurs de prédisposition. De plus, l'exposition à des agents tératogènes, tels que des médicaments ou des traumatismes, peut également entraîner des défauts de développement similaires à ceux provoqués par des accidents génétiques, en interagissant avec l'activité des gènes du développement. On peut citer à titre d'exemples les anomalies chromosomiques, touchant des gènes codant les nombreux facteurs de transcription qui régissent sur l'oculogenèse.

Ces questions mettent en relief tout l'intérêt des études en Biologie du Développement qui reposent sur la mise au point et l'exploitation de modèles expérimentaux, où l'analyse des relations épistatiques entre les gènes permet, d'une part, l'élaboration des réseaux génétiques impliqués, l'étude de leur cinétique d'action, et d'autre part, l'identification de facteurs de convergence comme cible thérapeutique potentielle.

—

**Figure 1:**

(A) Vue dorsale en stéréomicroscopie d'un embryon humain au jour gestationnel (J)18 après fécondation. La partie caudale, en bas, est toujours en cours de gastrulation pendant que la partie rostrale entame déjà la neurulation. (B) Vue dorsale d'un embryon du même stade par microscopie électronique à balayage. Zone agrandi en (C) indiqué. (C) A l'extrémité la plus rostrale de la ligne primitive, des cellules de l'épiblaste se détachent et migrent en tant que mésenchyme lâche dans le sens des flèches entre épiblaste et hypoblaste mais aussi vers la tête, déposant progressivement tout le mésoderme suivant une distribution rostrocaudale. (D) J21. La gouttière neurale est ouverte dans la région céphalique, vers le haut. (E) Vue dorsale, tête à gauche, J21. Le mésoderme s'organise en paires de blocs épithéliales, les somites, de part et d'autre de la gouttière neurale, à l'origine des futures structures segmentées du corps, à savoir les vertèbres, côtes, et muscles. (F) Au début de la 4$^e$ semaine de gestation, le tube neural se ferme et se détache sous l'ectoderme dorsal, alors que les somites continuent à se former en l'accompagnant de rostral en caudal. (G) En vue frontale de la face présomptive, la fermeture du tube neural n'est pas encore achevée, laissant apercevoir le prosencéphale. L'ébauche du cœur se développe à proximité du cerveau antérieur avant de s'en éloigner par la formation des arcs pharyngés au cours de la semaine qui suit. (H) La partie restant ouvert du tube neural en rostral est le neuropore antérieur, J26. (I) Le neuropore postérieur est encore ouvert au cours de la 4$^e$ semaine de gestation. Ant, antérieur ; Di, diencéphale ; Mes, mésencéphale ; Post, postérieur ; Rh, rhombencéphale ; So, somite. Images de Dr. K. K. Sulik, avec permission.

**Figure 2:**

Les cellules de la crête neurale (CCN), lorsqu'elles se détachent des bourrelets neuraux, débutent d'importantes migrations qui les conduisent à essaimer dans tout l'embryon. (A) Schémas en vue transversale au niveau céphalique ; CCN (rouges) initialement au sein des bourrelets neuraux puis, en tant que mésenchyme qui migre à distance du tube neural. (B) Photomicrographies en vue dorsale de la région céphalique d'embryons de poulet autour de 30-35 heures d'incubation, avec 6 à 8 paires de somites (ss) et rostral vers le haut. Les CCN sont marquées en fluorescence rouge. Elles se détachent du tube neural dorsal en même temps que les vésicules optiques s'évaginent progressivement du diencéphale latéral, sous cette chape de mésenchyme. (C) Embryon humain, vue dorsale en microscopie électronique à balayage, vers le $24^e$ jour de gestation. Tube neural indiqué en rouge. Les dérivés cellulaires des CCN proviennent de différents niveaux le long du tube neural, et sont plus nombreux à partir des CCN céphaliques que troncales. Certains dérivés, tels les tissus structuraux issus du mésectoderme comme la sclère ou les os de la face, sont produits uniquement par des CCN céphaliques ; d'autres, tels certains dérivés endocriniens ou le système nerveux (S.N.) entérique, ne proviennent que de régions très spécifiques, délimitées selon le niveau de l'axe rostrocaudal. A, B : images des auteurs ; C, photomicrographie de Dr. K.K. Sulik, avec permission.

**Figure 3 :**

Les régions du cerveau se différencient avant la fin du premier mois de grossesse. (A) Les évaginations optiques (flèches) sont visibles dans le neuroépithelium dès le $22^e$ jour (J) de gestation. (B) Les sous-divisions anatomiques du cerveau sont observées dès la $6^e$ semaine de gestation, ici dans un schéma de la vue latérale gauche. La vésicule optique gauche figure en blanc. (C) Vue fronto-latérale d'un embryon humain a J22 en microscopie électronique à balayage ; le diverticule optique est délimité en orange. (D, E) Schémas en coupe transversale des changements morphogénétiques du cerveau antérieur et champs optiques entre le $22^e$ (D) et $26^e$ (E) jours de gestation et la fermeture des bourrelets du diencéphale. A, C: Adapté de Dr. K. K. Sulik, avec son accord. B : Adapté de Wikimedia Commons. D, E : Adaptés d'Essentials of Human Embryology, William J. Larsen, avec l'accord d'Elsevier.

**Figure 4**

Induction du cristallin et morphogenèse de la cupule optique. (A) A la fin du 1$^{er}$ mois de gestation, plusieurs mouvements tissulaires se passent simultanément dans la vésicule optique (flèche). (B) Ces illustrations de coupes histologiques chez deux embryons humains du même stade montrent la fugacité des étapes qui marquent le passage de l'induction du placode cristallinien et du disque rétinienne à la démarcation de la tige optique et l'oblitération de l'espace sous-rétinienne (ESR). (C) La cupule et la tige optique proviennent du neuro-ectoderme. Sous l'ectoderme, des cellules mésenchymateuses (M), largement d'origine crête neurale, entourent la cupule optique. (D) Vue en coupe partielle montrant la fissure optique sur la face inférieure de la tige, qui résulte du contact de la rétine neuro-sensorielle (n.s.) avec la couche du futur épithélium pigmentaire rétinien (EPR). Une animation de ce processus peut être visionnée à ce lien : http://www.nature.com/nrn/journal/v8/n12/extref/nrn2283-s1.swf

**Figure 5**

Fermeture de la tige optique et développement des vaisseaux hyaloïdes à la fin de la 6$^e$ semaine du développement (sd). (A) Entre les stades 13 et 18 de Carnegie, c'est-à-dire entre 5 et 8 sd chez l'embryon humain, plusieurs mouvements morphogénétiques se coordonnent dans la région optique pendant la transformation de la vésicule à une structure oculaire plus complexe. (B) Œil microdisséqué d'embryon humain vers J37, montrant la fissure optique (flèche) coté ventral et la pigmentation de l'épithélium pigmentaire rétinien par transparence. (C) Coupe histologique dans le plan délimité en (B). La constriction de la tige optique et de l'espace sous-rétinienne sont évidentes, ainsi que le détachement de la vésicule cristallinienne. (D) Coupe histologique dans le même plan, vers J42. Des cellules mésenchymateuses se trouvent entre l'ectoderme et le cristallin (flèche pleine) ainsi qu'en accompagnement (étoile) des vaisseaux hyaloïdes (flèche vide) dans la fissure optique.

**Figure 6**

Maturation du cristallin et de l'angle irido-cornéen. (A) Illustration de 1907 de l'anatomie de l'angle irido-cornéen humain, autour de 19 semaines de développement (sd). (B) Photomicrographie de 1989 montrant les structures analogues chez un fœtus vers 22 sd. (C-E) Photomicrographies en immunofluorescence traçant le devenir de la majorité des cellules de la crête neurale (CCN) chez des souris transgéniques. Les CCN sont en vert, les noyaux en (C) sont en bleu et les érythrocytes en (D) sont en rouge. A 10,5 jours (J) sur 21 de gestation de la souris, la forme de l'œil est similaire à l'œil humain vers la fin de 6 sd. Les CCN investissent l'espace entre ectoderme et cristallin pour donner la cornée, ainsi que l'espace derrière le cristallin pour participer au tissu conjonctif des vaisseaux hyaloïdes. (D) Les CCN persistent en tant que péricytes de l'ensemble des vaisseaux sanguins de l'œil à la naissance. (E) L'angle irido-cornéen présente un fort contingent de CCN, notamment dans le stroma cornéen, à l'exception des processus ciliaires, la rétine neuro-sensorielle et le cristallin. AT, ébauche trabéculaire; CM, muscle extra-oculaire ; cs, canal de Schlemm; C, cornée ; ec/EC: endothélium cornéen ; ICA, angle irido-cornéen ; L, cristallin ("lens crystallina") ; mc, muscle ciliaire ; pc, processus ciliaires ; R, rétine ; RBC, érythrocyte ; sp, sphincter pupillae ; S, stroma cornéen.

**Figure 7**

La maturation du cristallin fait en sorte que les premières cellules, cuboïdes, ne persistent que sur la face antérieure alors que les fibres secondaires, longitudinales, croissent pour combler la cavité (A). (B) La chambre antérieure à 8 semaines de développement (sd) par microscopie électronique à balayage ; le stroma de la cornée (C) est présent ainsi que la membrane pupillaire (flèche) qui recouvre le cristallin. (C) L'intérieur du cristallin à 7 sd montre le noyau embryonnaire (N) de la structure et l'épithélium fin, antérieurement (flèche). (D) Sous le cristallin et la membrane pupillaire, on distingue la chambre antérieure (*) et les vaisseaux sanguins. (E) A plus fort grossissement, la marge antérieure de la cupule optique (O) est visible ainsi que des fentes dans la région limbique de la cornée (C) qui se rejoignent pour former le canal de Schlemm. La flèche indique le bord de la membrane pupillaire. (F) Vers 13 sd, les processus ciliaires (flèche) se forment dans l'iris postérieur exposant ainsi la trame trabéculaire. Schéma par T. C. Hengst; photos par Dr. K. K. Sulik, avec permission.

**Figure 8**

Des mutations héritées dans de gènes codant de nombreuses protéines de la région ciliaire du segment externe sont responsables de rétinites pigmentaires isolés et syndromiques : associés ou non à l'amaurose de Leber, aux syndromes d'Usher, de Bardet-Biedl et de Joubert. (A) Immunofluorescence anti-rhodopsin (bâtonnets, verts) et anti-calbindin (cônes et cellules horizontales, rouge). (B) Microscopie électronique à transmission pour montrer par des points de déposition de particules d'or au pont ciliaire, la localisation de myosin 7A, dont la mutation du gène est responsable du syndrome d'Usher type B. A : photo par Dr. Robert Fariss, National Eye Institute, NIH, avec permission. B : modifié de Liu, X. et al., 1997: Cell Motility and the Cytoskeleton, 37(3), pp.240–252.

**Figure 9**

De l'extérieur vers l'intérieur de la rétine, plusieurs couches histologiques sont identifiables chez la souris adulte comme chez l'humain. L'épithélium pigmentaire rétinien (EPR); la couche des segments externes (SE) des photorécepteurs, en contact avec l'EPR ; le segment interne au contact avec la membrane limitante externe (mli) ; la couche nucléaire externe (CNE) ; la couche plexiforme externe (CPE); la couche nucléaire interne (CNI); la couche plexiforme interne (CPI) ; la couche des corps des cellules ganglionnaires (CG) ; la couche des axones ganglionnaires (AG) qui convergent vers le nerf optique ; la membrane limitante interne (mli). Photographie des auteurs.

**Figure 10**

Interactions entre la couche interne vasculaire de la choroïde, l'épithélium pigmentaire (EPR) et les photorécepteurs de la rétine. En haut, microscopie électronique à balayage des trois couches vasculaires de la choroïde dont la plus interne, le choriocapillaris, est séparée par une lame basale du contact direct avec l'EPR. Illustration adaptée de Zhang, H.R., 1994. Scanning electron-microscopic study of corrosion casts on retinal and choroidal angioarchitecture in man and animals. Progress in Retinal and Eye Research, 13(1), pp.243–270. © 1994, avec l'accord d'Elsevier ; et de Strauss O (2005) The retinal pigment epithelium in visual function. Physiol Rev 85:845-81. m, melanosome ; MB, membrane de Bruch ; mv, microvillosités ; PEDF, Pigment epithelium-derived factor (serpin F1) ; SE, segment externe ; VEGF, vascular endothelial growth factor.

**Figure 11**

La vascularisation de la rétine en profondeur commence par le plexus rétinien intérieur qui se développe à partir des vaisseaux hyaloïdes à la fin de 8 semaines de développement (sd). (A, B) Schémas en vue latérale montrent les ramifications de l'artère hyaloïde, visibles également en (C) sur coupe histologique avec coloration à l'éosine. (D) Vers 25 sd, au niveau du fovea. (E, F) Immunohistochimie contre CD34, ce qui met en évidence les cellules endothéliales du plexus rétinien intérieur et extérieur en formation, respectivement, à 26 sd. Une artériole (a) est également visible. A, B : Reproduit avec permission de [27] ; C : Hill, M.A. (2016) « Embryology » *Stage 22 image 008-eye.jpg*. Récupéré le 30 juin 2016, de https://embryology.med.unsw.edu.au/embryology/index.php/File:Stage_22_image_008-eye.jpg ; D-F : Reproduit avec permission de [13].

**Figure 12**

Maturation de la cornée et des paupières. (A) La cornée reçoit un important contingent de fibres sensoriels provenant du ganglion trigéminal, tout d'abord aux bords périphériques de la cornée, mais qui s'étend de façon centripète entre le $3^e$ et $5^e$ mois. (B) Le processus de maturation de la cornée est initié dans les couches les plus profondes de la cornée, puis progresse plus superficiellement. (C) A partir de 6 semaines de développement (sd), les paupières se forment à partir d'une expansion de l'ectoderme. (D) Elles se ferment vers 9-10 sd. (E) Les paupières restent soudées pendant plusieurs mois. (F) Début de formation des follicules ciliaires. (G) Développement des follicules des cils se poursuit (a) alors que des invaginations postérieures signalent le début de la formation des glandes de Meibomius (b) et le muscle orbicularis se différencie au centre. (H) La progression de la différentiation ciliaire et musculaire se poursuit. La disjonction des paupières débute antérieurement vers 20 sd. F à H de Hamming, N., 1983. Anatomy and embryology of the eyelids: a review with special reference to the development of divided nevi. Pediatr. Dermatol. 1, 51–8, avec l'accord de Wiley.

# Bibliographie